\documentclass[twocolumn,showpacs,preprintnumbers,amsmath,amssymb]{revtex4}

\usepackage{graphicx}

\begin{document}
\title{Transient Rectification of Brownian Diffusion with
 Asymmetric Initial Distribution}
\author{A.V. Plyukhin and A.M. Froese}
 \affiliation{ Department of Physics and Engineering Physics,
 University of Saskatchewan, Saskatoon, SK S7N 5E2, Canada 
}

\date{\today}

\begin{abstract}
In an ensemble of non-interacting Brownian particles, a 
finite systematic average velocity may temporarily develop, 
even if it is zero initially. The effect originates from a small
nonlinear correction to the dissipative force,
causing the equation for the first moment of velocity to couple to 
moments of higher order.
The effect may be relevant when a
complex system dissociates in a viscous medium with conservation of momentum.
\end{abstract}


\maketitle

\section{Introduction}
Stochastic processes with nonlinear dissipation 
are ubiquitous
and challenging to describe theoretically.
Mathematical difficulties related to the 
nonlinearity of a corresponding stochastic differential 
equation are only part of
the problem. A more subtle challenge is to establish 
fluctuation-dissipation relations which, in contrast to linear
processes, cannot be phenomenologically justified~\cite{Kampen_book}. 
Instead, a truly dynamical approach is usually needed when the 
dissipation force and 
statistical properties of the noise are deduced directly from
underlying dynamics, rather than postulated {\it ad hoc}.
Conventional assumptions of 
a phenomenological approach in the context of nonlinear response
may be misleading. For instance, the assumption of Gaussian random 
force in the Langevin
equation leads to the Fokker-Planck equation of second order,
regardless of whether the dissipation force is linear or not.
On the other hand, a kinetic approach leads to the second-order 
Fokker-Planck equation for a Brownian particle only in the lowest
order of a perturbation technique, while in general 
the equation involves derivatives of order higher 
than two~\cite{Kampen_book,Kampen, Kampen2,PhysA}.

Nonlinear stochastic processes are usually associated with
far-from-equilibrium dynamics. If a system is close to 
equilibrium, nonlinear dissipation usually appears 
as small corrections to the dominating linear friction
and in many cases may be safely neglected. 
However, under certain circumstances, 
the contribution of linear terms may vanish identically or be 
strongly reduced. Then nonlinear dissipative effects
come into the limelight and give rise to a variety of new physical effects. 

An example, which has received particular attention in recent years, 
is the rectification of thermal fluctuations in 
the so-called adiabatic piston problem~\cite{Gruber}. 
The problem concerns Brownian motion
of a piston which separates a gas-filled
cylinder into two compartments with different temperatures and gas densities. 
If the pressure on both sides of the piston is the same, the linear
theory predicts zero average velocity of the piston, while the correct
result is that the piston acquires a systematic average speed in the
direction of the compartment with higher temperature. 
The effect may
be readily explained using the Langevin equation with a small nonlinear
correction, quadratic in the piston's velocity, 
to the dissipative force~\cite{PS_piston}. 
Some other effects related to the nonlinear dissipation are discussed 
in~\cite{Gelin}. 

In the adiabatic piston problem the fluctuation-induced drift originates from
nonequilibrium and asymmetry.
The current point of view is that these two ingredients are
necessary in general for rectification of thermal fluctuations, i.e.
for the physical realization of Maxwell's demon.
Asymmetry may be introduced by surroundings, as in the adiabatic
piston problem, or by the geometry of the
Brownian particle itself~\cite{Broeck1,Broeck2}. In this paper, our concern is
a transient rectification effect originating 
from asymmetric initial conditions. 
\begin{figure}
\hfill{}\includegraphics[scale=0.5]{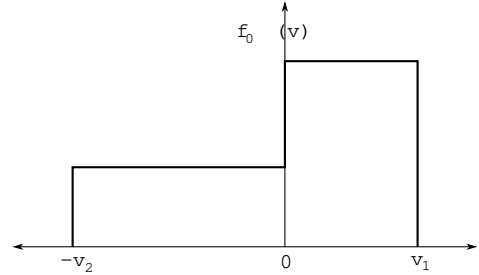}\hfill{}
\caption{ 
Initial velocity distribution for an ensemble of Brownian particles,
discussed in the paper.
The widths and heights of the distribution's wings are chosen
so that the average initial velocity $\langle V\rangle$
of the ensemble is zero, but the higher moments $\langle V^n\rangle$
are finite.}
\end{figure}

\section{The problem}
Consider an ensemble of non-interacting 
Brownian particles diffusing in one dimension.
The particles are identical but may have different initial velocities. 
Suppose the distribution of initial velocities $f_0(V)$ is similar
to Fig. 1: asymmetric but in such a way that the average 
initial velocity of the ensemble is zero, 
\begin{eqnarray}
\langle V(0)\rangle=\int dV f_0(V) V=0.
\label{symmetry}
\end{eqnarray}
The question is whether $\langle V(t)\rangle$ for
later time $t>0$ is positive, negative or zero?

Contrary to its apparent simplicity, the question
requires going beyond the standard theory of Brownian
motion based on the linear Langevin equation and the corresponding
second-order Fokker-Planck equation.
Both approaches give the linear relaxation law
$\partial_t\langle V(t)\rangle=-\gamma\langle V(t)\rangle$, and therefore 
predict that if the average velocity $\langle V(t)\rangle$
is zero initially, it remains so later on. The prediction is incorrect
as one can see from the result of numerical experiment presented in
Fig. 2 and Fig. 3. On the time scale of order $\tau=1/\gamma$, 
a finite average velocity develops in the direction 
corresponding to the higher, narrower
wing of the initial distribution, the right wing in Fig. 1.
The particles moving to the right,
have lower average initial speeds but are more numerous and 
give a larger contribution to $\langle V(t)\rangle$
than the particles moving to the left. 
The victory of the larger team of slower runners does not last very
long: after
reaching a peak at roughly one half of $\tau$, 
the function $\langle V(t)\rangle$ decays
exponentially with the characteristic time of order $\tau$.
Yet, this transient time 
may be sufficiently long to cause measurable physical consequences. 

\begin{figure}
\hfill{}\includegraphics[scale=0.6,angle=-90]{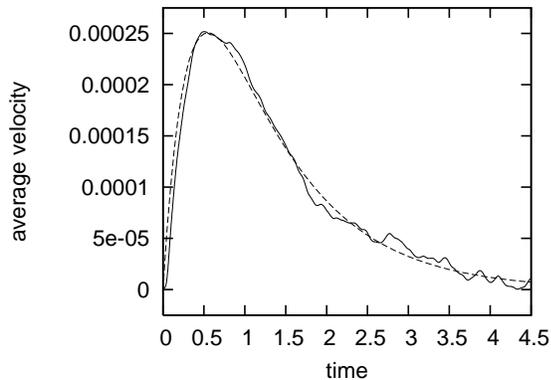}\hfill{}
\caption{ 
Simulation (solid) and theoretical (dashed) curves for 
the time dependence of the average 
velocity $\langle V(t)\rangle$ of an
ensemble with an initial distribution similar to Fig. 1.
The molecule-particle mass ratio parameter is $\lambda=\sqrt{m/M}=0.1$.
The widths of the distribution wings are $V_1=1/4$ and
$V_2=1/2$. Velocity is in units $v_{th}=\sqrt{kT/m}$ and
time is in units $\tau=(\lambda^2\gamma_0)^{-1}$.
}
\end{figure}

\begin{figure}
\hfill{}\includegraphics[scale=0.6,angle=-90]{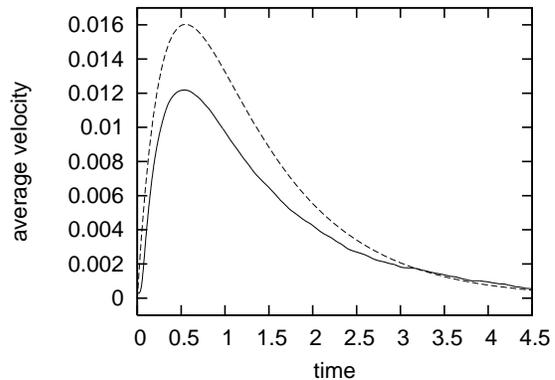}\hfill{}
\caption{ Same as for Fig. 2, but for an initial velocity distribution
with widths $V_1=1$ and $V_2=2$ (in units $v_{th}$)
for the right and left wings, respectively.
The corresponding ensemble is far from equilibrium,
$V_1, V_2>\lambda$. The theoretical (dashed) curve,
given by Eq. (\ref{SOL}) overestimates the result of
the simulation (solid curve).
}
\end{figure}

The problem may be considered as an idealized model of the dissociation
of a complex system in a viscous medium. 
Among possible relevant fields are 
the Coulomb fragmentation of multiply-charged clusters and 
droplets~\cite{Calvo} and processes involving 
fragmentation of complex molecular aggregates, such as protein-ligand 
dissociation~\cite{Solo}.
If the system is initially at rest and all dissociated fragments
have the same mass, Eq. (\ref{symmetry}) is just the condition of
conservation of total momentum.
For a system in vacuum, the
speed of the center of mass of fragments remains zero after
dissociation. 
However if dissociation happens in a viscous medium, 
the average velocity is temporarily finite, and 
the center of mass changes position even if the fragments
are identical and have the same diffusion coefficients.

To account for this transient rectification effect, one has to take
into account that the equation for the first moment of the
velocity $\partial_t\langle V(t)\rangle=-\gamma\langle V(t)\rangle$
is closed only in lowest order in the small parameter $\lambda^2=m/M$,
the mass ratio of a molecule ($m$) to a Brownian particle ($M$).
At higher orders in $\lambda$, the first moment 
$\langle V(t)\rangle$ is coupled to the moments of higher orders 
$\langle V^n(t)\rangle$. If initially the first moment is zero, 
but the higher moments are finite, as for the initial distribution
in Fig. 1, then $\langle V(t)\rangle\ne 0$ for $t>0$.
To describe the problem quantitatively, one may adopt the approach
based on either the Langevin equation for $V(t)$ or the Fokker-Planck
equation for the distribution function $f(V,t)$. 
In what follows,
we discuss both approaches and outline details of the numerical
simulations presented in Fig. 2 and Fig. 3.

\section{Theory: Langevin equation}
The microscopic derivation of the Langevin
equation beyond the lowest order in $\lambda=\sqrt{m/M}$ was
discussed recently in detail in \cite{PS_piston_basic}.
Here we outline the results and apply them to our problem.
An appropriate perturbation technique is guided by
anticipation that the velocity $V$ of a Brownian particle is typically
about $\lambda$ times that of a molecule of the surrounding
bath. 
This suggests working with the scaled velocity of the
particle $v=\lambda^{-1}V$, 
which is expected to be of the same order as the thermal velocity of 
molecules $v_{th}$, 
\begin{eqnarray}
v=\lambda^{-1}V\sim v_{th}=\sqrt{\frac{kT}{m}}.
\end{eqnarray}
The microscopic equation of motion 
for the scaled velocity $v=V/\lambda$ (or for the scaled momentum
$p=mv=\lambda MV$) involves
the small parameter $\lambda$ explicitly, and therefore 
is convenient for a perturbation analysis. The equation is coupled
with bath degrees of freedom which may be ``projected out'' with an
appropriate projection operator technique~\cite{Mazur,PS_piston_basic}. 
As a result, to lowest
order in $\lambda$, one obtains the conventional linear Langevin equation
\begin{eqnarray}
\dot{v}(t)=-\lambda^2\gamma_0\, v(t)+\frac{\lambda}{m}\, F_0(t),
\label{LE1}
\end{eqnarray}
where the zero-centered fluctuating force $F_0(t)$ is related to
the dissipation constant 
$\gamma_0$ through the 
fluctuation-dissipation relation 
\begin{eqnarray}
\gamma_0=\frac{1}{mkT}\,\int_0^\infty dt\,\langle F_0(0)F_0(t)\rangle.
\label{FDT}
\end{eqnarray}
The linear Langevin 
equation (\ref{LE1}) leads to the following equations for the velocity
moments~\cite{LE_book}
\begin{eqnarray}
\frac{d\langle v^n\rangle}{dt}=-\lambda^2\, n\,\gamma_0\, \langle v^n\rangle
+\lambda^2 \,n\,(n-1)\,\gamma_0\,v^{-2}_{th} \langle v^{n-2}\rangle.
\label{moments}
\end{eqnarray}
As discussed, 
these equations, obtained in lowest order in $\lambda^2$, 
are not sufficient for our purpose: 
the closed equation for the first moment
$\partial_t\langle v\rangle=-\lambda^2\gamma_0\langle v\rangle$
clearly cannot account for the behavior presented in Fig. 2.

The next approximation for the Langevin equation involves a
correction of order 
$\lambda^4$ and, for a homogeneous bath, has the form~\cite{PS_piston_basic}
\begin{eqnarray}
\dot{v}(t)=-\lambda^2\,\gamma_1\, v(t)-\lambda^4\, \gamma_2\, v^3(t)+
\frac{\lambda}{m}\,F(t).
\label{LE2}
\end{eqnarray}
Besides the presence of the nonlinear dissipative term 
$-\lambda^4\gamma_2\, v^3$, this equation differs from the linear one
(\ref{LE1}) by a higher order correction to the linear damping, and the
fluctuating force
\begin{eqnarray}
\gamma_1=\gamma_0+O(\lambda^2),\,\,\,F(t)=F_0(t)+O(\lambda). 
\label{aux1}
\end{eqnarray}
The explicit form of these corrections is not necessary for the purpose
of this paper.
The fluctuation-dissipation relation
for the nonlinear dissipation coefficient 
$\gamma_2$ involves rather complicated correlation
functions~\cite{PS_piston_basic}, and 
to the best of our knowledge, cannot be established
phenomenologically. This is in contrast to 
the conventional fluctuation-dissipation relation (\ref{FDT}) for the
linear dissipation coefficient $\gamma_0$ which 
can be obtained  using the prediction of equilibrium
statistics $\langle v^2(t)\rangle\to kT/m$ in the long time limit. 

Since the fluctuating force is zero-centered to any order in $\lambda$,
it follows from Eq. (\ref{LE2}) that to order $\lambda^4$ the first moment
is coupled to the third one, 
\begin{eqnarray}
\frac{d}{dt}\langle v\rangle=-\lambda^2\gamma_1 \,\langle v\rangle
-\lambda^4 \gamma_2 \,\langle v^3\rangle.
\label{eq1}
\end{eqnarray} 
One has to substitute 
here $\langle v^3(t)\rangle$
obtained in the lowest order in $\lambda$ which according to
(\ref{moments})
satisfies the equation
\begin{eqnarray}
\frac{d}{dt}\langle v^3\rangle=-3\lambda^2\gamma_0\, \langle v^3\rangle
+6\lambda^2\gamma_0\, v_{th}^{-2}\,\langle v\rangle.
\label{eq2}
\end{eqnarray}
Our interest is the solution of Eqs. (\ref{eq1}) and (\ref{eq2}) with the
initial conditions 
\begin{eqnarray}
\langle v(0)\rangle=0, \,\,\,\,\langle v^3(0)\rangle\ne 0.
\end{eqnarray}
Clearly, in this case $\langle v(t)\rangle\sim \lambda^2$, so that 
the last term in the Eq. (\ref{eq2}) can be neglected. Then, to 
order $\lambda^2$, the third moment decays exponentially 
$\langle v^3(t)\rangle =\langle v^3(0)\rangle\,e^{-3\lambda^2\gamma_0t}$.
Substituting this into Eq. (\ref{eq1}) and recalling that
$\gamma_1=\gamma_0+O(\lambda^2)$, one obtains
\begin{eqnarray}
\langle v(t)\rangle=
-\lambda^2\frac{\gamma_2}{2\gamma_0}\,\langle v^3(0)\rangle\, 
e^{-\lambda^2\gamma_0 t}(1-e^{-2\lambda^2\gamma_0 t}).
\label{sol}
\end{eqnarray} 
Recall also that $v$ is the scaled velocity, $v=V/\lambda$. For the true
velocity $V$ the result formally does not involve the small factor
$\lambda^2$,
\begin{eqnarray}
\langle V(t)\rangle=
-\frac{\gamma_2}{2\gamma_0}\,\langle V^3(0)\rangle\, 
e^{-\lambda^2\gamma_0 t}(1-e^{-2\lambda^2\gamma_0 t}).
\label{Sol}
\end{eqnarray}
However, one should keep in mind that the whole procedure applied above 
implies that $V\sim\lambda\, v_{th}$. This puts a constraint on the
the width $\Delta $ of the initial distribution $f_0(V)$,
\begin{eqnarray}
\Delta<\lambda v_{th}.
\label{constraint}
\end{eqnarray}
Under this constraint $\langle V^3(0)\rangle$ is small and 
cannot exceed order $\lambda^3\, v_{th}^3$. 

For a far-from-equilibrium
ensemble the above theory, strictly speaking,
is not applicable. 
Yet, as one observes from Fig. 3, Eq. (\ref{Sol}) predicts
qualitatively correct behavior also for a ``hot'' initial distribution 
with $\Delta\sim v_{th}$.
In these cases the first moment given by Eq. (\ref{Sol}) 
is not small, $\langle V(t)\rangle\sim \lambda^0$.

According to the result (\ref{Sol}), 
the first moment $\langle V(t)\rangle$ reaches the maximum
at time $t_0=(\ln{3}/2)\tau\approx 0.55\,\tau$ where
$\tau=\lambda^{-2}\gamma_0^{-1}$, which is seen in Fig. 2
to be in agreement with numerical simulation.
To make more qualitative predictions, 
one needs an explicit
expression for the ratio of the dissipative coefficients
$\gamma_2/\gamma_0$, which is the prefactor in Eq. (\ref{Sol}).
Since a general result for this ratio is unknown, in 
the rest of the paper 
we discuss a specific model of Brownian motion - the Rayleigh model - 
for which our numerical experiment is carried out, and for which 
analytical results are available. 

In the original Rayleigh model~\cite{Kampen,Kampen2,PhysA}, a heavy 
Brownian particle moves in one dimension interacting with bath
molecules through instantaneous elastic collisions, while molecules do
not interact with one another at all. For this model the Fokker-Planck equation
for the distribution function $f(V,t)$ 
can be readily obtained, as will be discussed
in the next section. However, due to the singular character of the hard-wall
potential, the derivation of a nonlinear Langevin equation for the original
Rayleigh model is not quite 
straightforward. One may instead to work with a generalized Rayleigh
model where the particle interacts with molecules through a continuous 
repulsive potential. For a low density of bath molecules (when multiple
collision are negligible) and for the
time scale longer than the collision time $\tau_c$, the original and
generalized models are expected to give the same results. 
Using the generalized Rayleigh model, one obtains the following  
explicit expressions for the dissipative 
coefficients~\cite{PS_piston_basic} 
\begin{eqnarray}
\gamma_0=\frac{8}{\sqrt{2\pi}}\, n\,S\,v_{th} ,\,\,\,\,
\gamma_2=\frac{4}{3\sqrt{2\pi}}\,n\,S\,v_{th}^{-1}. 
\label{gammas}
\end{eqnarray} 
Here $n$ is the concentration of molecules, $S$
is the particle's cross-section, and $v_{th}=\sqrt{kT/m}$
is the thermal velocity of molecules in the bath.
It is tempting to assume that the relation
\begin{eqnarray}
\frac{\gamma_2}{\gamma_0}=\frac{1}{6}\,v_{th}^{-2}=\frac{m}{6\,kT},
\label{ratio}
\end{eqnarray}
which follows from (\ref{gammas}), 
is in fact general but we leave this conjecture for further studies.
Substituting (\ref{ratio}) into Eq. (\ref{Sol}), one finally obtains
\begin{eqnarray}
\langle V(t)\rangle=
-\frac{m}{12\,kT}\,\langle V^3(0)\rangle\, 
e^{-\lambda^2\gamma_0 t}(1-e^{-2\lambda^2\gamma_0 t}).
\label{SOL}
\end{eqnarray}
Subsequently, 
the average displacement of the ensemble is 
\begin{eqnarray}
\langle X \rangle=\int_0^\infty dt\, \langle V(t)\rangle=
\frac{1}{18}\,
\frac{1}{\gamma_0\lambda^2v_{th}^{2}}\,\langle V^3(0)\rangle.
\end{eqnarray}

The result (\ref{SOL}) for $\langle V(t)\rangle$,
presented in Fig. 2 by dashed lines,
is in good agreement with numerical simulation as long as the constraint
(\ref{constraint}) on the initial distribution is satisfied. 
Before discussing details of the simulation, let us derive the 
results using the language of the Fokker-Planck equation.

\section{Theory: Fokker-Planck equation}
For the original Rayleigh model, which involves only binary-particle
molecule collisions, 
the Fokker-Planck equation can be readily obtained using the
Kramers-Moyal expansion of the master 
equation~\cite{Kampen,Kampen2,PhysA,Broeck1}.
To order $\lambda^2$, the equation has a familiar form 
\begin{eqnarray}
\frac{\partial f(v,t)}{\partial t}=\lambda^2 \gamma_0\, D_2 f(v,t),
\label{FPE1}
\end{eqnarray}
where the second order differential operator $D_2$ reads 
\begin{eqnarray}
D_2=\frac{\partial}{\partial v} v+
v_{th}^2 \, \frac{\partial^2}{\partial v^2}
\label{D2}
\end{eqnarray}
and $\gamma_0$ is given by (\ref{gammas}).
This equation corresponds to the linear Langevin equation (\ref{LE1})
and produces Eq. (\ref{moments}) for the moments $\langle
v^n(t)\rangle$ to order $\lambda^2$. 
The equation of order $\lambda^4$ has the 
form~\cite{Kampen,PhysA}
\begin{eqnarray}
\frac{\partial f(v,t)}{\partial t}=\lambda^2\gamma_0\, D_2 f(v,t)+
\lambda^4\gamma_0\, D_4 f(v,t).
\label{FPE2}
\end{eqnarray}
where the forth-order differential operator $D_4$ reads
\begin{eqnarray}
D_4&=&
-\frac{\partial}{\partial v} v
+\frac{1}{6}\,v_{th}^{-2}\frac{\partial}{\partial v} v^3
-2\,v_{th}^2\frac{\partial^2}{\partial v^2}\label{D4}\\
&&+\frac{3}{2}\,\frac{\partial^2}{\partial v^2} v^2
+\frac{8}{3}\,v_{th}^2\frac{\partial^3}{\partial v^3} v
+\frac{4}{3}\,v_{th}^4\frac{\partial^4}{\partial v^4}\nonumber.
\end{eqnarray} 
For the first moment, Eq. (\ref{FPE2}) gives the following equation
\begin{eqnarray}
\frac{d}{dt}\langle v\rangle=-\lambda^2\gamma_0(1-\lambda^2)\,\langle
v\rangle
-\frac{1}{6}\,\lambda^4\gamma_0\,v_{th}^{-2}\langle v^3\rangle.
\end{eqnarray}
Recalling the relations (\ref{aux1}) and (\ref{ratio}), 
one observes that the above
equation is equivalent to Eq. (\ref{eq1}) derived from the
nonlinear Langevin equation. Therefore, the 
Fokker-Planck equation (\ref{FPE2}) gives the same results as the nonlinear
Langevin equation~(\ref{LE2}). Note, however, that the Langevin 
equation~(\ref{LE2}) is derived directly from the Liouville
equation~\cite{PS_piston_basic} and is more general than the
Fokker-Planck equation (\ref{FPE2}), which is obtained under the assumption
of binary particle-molecule collisions.

\section{Simulation}
In our molecular dynamics
simulation, we use the generalized Rayleigh model in which 
the Brownian particle moves in one dimension interacting with 
molecules through a finite-range repulsive parabolic potential,
while molecules do not interact with one another.
In this model, discussed in detail in \cite{PS_piston_basic},
the particle-molecule collision time $\tau_c$ is finite and does not
depend on the velocity of the molecule. A characteristic parameter of the
model is $N=nSv_{th}\tau_c$, which is an average number of molecules
simultaneously interacting with the particle. 
In simulation, the linear molecular density $nS$ was chosen to make 
$N$ of order 1. In this case, multiple particle-molecule
collisions are rare, and one can
expect that the result should be close to that for the
original Rayleigh model with instantaneous binary collisions.

To mimic unbounded diffusion of a particle,
we have used two sources of molecules located far
from the particle that generate a bath
with a Maxwellian velocity distribution and a constant density.
The first condition is easily accommodated
by selecting incoming molecule velocities from the Boltzmann distribution
\begin{eqnarray}
\phi(v)=\frac{nSv}{v_{th}\sqrt{2\pi}}\exp\left(\frac{-v^{2}}
     {2v_{th}^{2}}\right),
\end{eqnarray}
while controlling the rate of molecule generation with a Poisson process
is one possibility that is consistent with the second condition.  With such a velocity distribution, the
total flux at each source is $\Phi=\int_0^\infty\phi(v)dv=nSv_{th}/\sqrt{2\pi}$.
The Poisson distribution for the period between molecule injections is
then $P(\tau_{in})=\exp(-\Phi t)$,
which will maintain an average linear density of $nS$ around the particle.

An ensemble of particles is emulated by performing
multiple runs, resetting the system between each run with the
new particle initial conditions
selected from the appropriate distribution functions, and averaging
the results of all runs together.
For a symmetric velocity distribution function $f_0(v)$, the simulation reproduced
familiar results of linear Brownian motion including the exponential decay 
of the velocity correlation function on a
time scale $t>\tau_c$ and deviation from
exponential form for $t<\tau_c$, which is in agreement with 
the theory developed in \cite{PS_piston_basic}.

Consider now an asymmetric initial distribution such as that shown in Fig. 1.
Let $x=V/v_{th}$ be the dimensionless velocity of the particle. Also let
$x_1, x_2$ be the widths and $c_1, c_2$ be the heights of
the right and left wings of the distribution $f_0(x)$, respectively. 
The conditions of
normalization $\int dx f_0(x)=1$ and of zero first moment $\int dx
 f_0(x)x=0$ give
\begin{eqnarray}
c_1x_1+c_2x_2=1,\,\,\,c_1x_1^2-c_2x_2^2=0
\end{eqnarray} 
and therefore,
\begin{eqnarray}
c_1=\frac{x_2}{x_1}\frac{1}{x_1+x_2},\,\,\,c_2=\frac{x_1}{x_2}
\frac{1}{x_1+x_2}.
\label{cs}
\end{eqnarray}
The theoretical prediction is given by Eq. (\ref{Sol}),
\begin{eqnarray}
\langle x(t)\rangle=
-\frac{1}{12}\,\langle x^3(0)\rangle\, 
e^{-t/\tau}(1-e^{-2t/\tau}),
\label{SOL2}
\end{eqnarray}
where $\tau=(\lambda^2\gamma_0)^{-1}$, 
and the initial third moment, according to (\ref{cs}), equals 
\begin{eqnarray}
\langle x^3(0)\rangle=\frac{x_1x_2}{4}(x_1-x_2).
\label{x3}
\end{eqnarray}

Recall that the theory outlined in previous sections applies under 
the close-to equilibrium constraint
(\ref{constraint}), which requires that $x_1$ and $x_2$ must be 
of order $\lambda$ or less. Note that for small $\lambda$, this condition is
not easy to satisfy in simulation. Since 
$\langle x(t)\rangle\sim \langle x^3(0)\rangle\le \lambda^3$,
one needs a very large number of runs (larger than $\lambda^{-6}$) 
to average out fluctuations and
find the function $\langle x(t)\rangle$ with reasonable precision. On the other
hand, 
a strongly non-equilibrium ensemble 
with the initial distribution widths
$x_1,x_2\sim 1$ is easier to simulate since in this case
$\langle x(t)\rangle\sim 1$, which requires a relatively small number of
runs.

The simulation has been performed for 
$\lambda=0.1$, $N=nSv_{th}\tau_c=1$, time step $\Delta
t=0.1\,\tau_c$, and various parameters of
the initial two-wing distribution $f_0(x)$ in Fig. 1. 
Time in Figs. 2 and 3 is given in units of velocity correlation time 
$\tau=(\lambda^2\gamma_0)^{-1}$ which, 
according to (\ref{gammas}), is related to 
the collision time $\tau_c$ by
$\tau_c/\tau=(8/\sqrt{2\pi})\lambda^2N$. 

Fig. 2 corresponds to the initial velocity distribution $f_0(x)$ with
left and right maximum velocities $x_1=1/4$ and $x_2=1/2$, respectively.
This is a close-to-equilibrium
ensemble, $x_1, x_2\sim \lambda$.
For this case, Eqs. (\ref{cs}) and (\ref{x3}) give
$c_1=8/3$, $c_2=2/3$, and 
$\langle x^3(0)\rangle=-1/128$.
As discussed
above, this case requires a large  number of runs to minimize
relative fluctuations. The presented plot (solid line)
is the average over about $5\times 10^7$ runs. 
Despite still visible fluctuations, 
the data and theoretical prediction (\ref{SOL})
are clearly in good agreement. 

Fig. 3 corresponds to the distribution with maximum velocities
$x_1=1$ and $x_2=2$. In this case, 
$c_1=2/3$, $c_2=1/6$, and $\langle x^3(0)\rangle=-0.5$.
The corresponding ensemble includes ``hot'' Brownian
particles with initial velocities $x>\lambda$ ($V>\lambda v_{th}$),
so that the major assumption of the theory is not satisfied.
It is not surprising then that in this case the 
theoretical prediction (\ref{SOL2}) distinctly overestimates
the simulation curve. 
Qualitative theory for 
a strongly non-equilibrium ensemble remains a challenge.

\begin{acknowledgments}
This work was supported by NSERC.
\end{acknowledgments}

\end{document}